\documentstyle[twoside,epsf]{article}

\catcode`\@=11
\long\def\@makefntext#1{
\protect\noindent \hbox to 3.2pt {\hskip-.9pt  
$^{{\eightrm\@thefnmark}}$\hfil}#1\hfill}		

\def\thefootnote{\fnsymbol{footnote}}
\def\@makefnmark{\hbox to 0pt{$^{\@thefnmark}$\hss}}	
	
\def\ps@myheadings{\let\@mkboth\@gobbletwo
\def\@oddhead{\hbox{}
\rightmark\hfil\eightrm\thepage}   
\def\@oddfoot{}\def\@evenhead{\eightrm\thepage\hfil
\leftmark\hbox{}}\def\@evenfoot{}
\def\sectionmark##1{}\def\subsectionmark##1{}}

\oddsidemargin=\evensidemargin
\renewcommand{\thefootnote}{\fnsymbol{footnote}}

\newcounter{sectionc}\newcounter{subsectionc}\newcounter{subsubsectionc}
\renewcommand{\section}[1] {\vspace{12pt}\addtocounter{sectionc}{1} 
\setcounter{subsectionc}{0}\setcounter{subsubsectionc}{0}\noindent 
	{\tenbf\thesectionc. #1}\par\vspace{5pt}}
\renewcommand{\subsection}[1] {\vspace{12pt}\addtocounter{subsectionc}{1} 
	\setcounter{subsubsectionc}{0}\noindent 
	{\bf\thesectionc.\thesubsectionc. {\kern1pt \bfit #1}}\par\vspace{5pt}}
\renewcommand{\subsubsection}[1] {\vspace{12pt}\addtocounter{subsubsectionc}{1}
	\noindent{\tenrm\thesectionc.\thesubsectionc.\thesubsubsectionc.
	{\kern1pt \tenit #1}}\par\vspace{5pt}}
\newcommand{\nonumsection}[1] {\vspace{12pt}\noindent{\tenbf #1}
	\par\vspace{5pt}}

\newcounter{appendixc}
\newcounter{subappendixc}[appendixc]
\newcounter{subsubappendixc}[subappendixc]
\renewcommand{\thesubappendixc}{\Alph{appendixc}.\arabic{subappendixc}}
\renewcommand{\thesubsubappendixc}
	{\Alph{appendixc}.\arabic{subappendixc}.\arabic{subsubappendixc}}

\renewcommand{\appendix}[1] {\vspace{12pt}
        \refstepcounter{appendixc}
        \setcounter{figure}{0}
        \setcounter{table}{0}
        \setcounter{lemma}{0}
        \setcounter{theorem}{0}
        \setcounter{corollary}{0}
        \setcounter{definition}{0}
        \setcounter{equation}{0}
        \renewcommand{\thefigure}{\Alph{appendixc}.\arabic{figure}}
        \renewcommand{\thetable}{\Alph{appendixc}.\arabic{table}}
        \renewcommand{\theappendixc}{\Alph{appendixc}}
        \renewcommand{\thelemma}{\Alph{appendixc}.\arabic{lemma}}
        \renewcommand{\thetheorem}{\Alph{appendixc}.\arabic{theorem}}
        \renewcommand{\thedefinition}{\Alph{appendixc}.\arabic{definition}}
        \renewcommand{\thecorollary}{\Alph{appendixc}.\arabic{corollary}}
        \renewcommand{\theequation}{\Alph{appendixc}.\arabic{equation}}
        \noindent{\tenbf Appendix \theappendixc #1}\par\vspace{5pt}}
\newcommand{\subappendix}[1] {\vspace{12pt}
        \refstepcounter{subappendixc}
        \noindent{\bf Appendix \thesubappendixc. {\kern1pt \bfit #1}}
	\par\vspace{5pt}}
\newcommand{\subsubappendix}[1] {\vspace{12pt}
        \refstepcounter{subsubappendixc}
        \noindent{\rm Appendix \thesubsubappendixc. {\kern1pt \tenit #1}}
	\par\vspace{5pt}}

\newcommand{\textlineskip}{\baselineskip=13pt}
\newcommand{\smalllineskip}{\baselineskip=10pt}

\def\eightcirc{
\begin{picture}(0,0)
\put(4.4,1.8){\circle{6.5}}
\end{picture}}
\def\eightcopyright{\eightcirc\kern2.7pt\hbox{\eightrm c}} 

\newcommand{\copyrightheading}[1]
	{\vspace*{-2.5cm}\smalllineskip{\flushleft
	{\footnotesize International Journal of Modern Physics C, #1}\\
	{\footnotesize $\eightcopyright$\, World Scientific Publishing
	 Company}\\
	 }}

\newcommand{\publisher}[2]{{\begin{center}\footnotesize\smalllineskip 
	Received #1\\
	Revised #2
	\end{center}
	}}

\def\abstracts#1#2#3{{
	\centering{\begin{minipage}{4.5in}\baselineskip=10pt\footnotesize
	\parindent=0pt #1\par 
	\parindent=15pt #2\par
	\parindent=15pt #3
	\end{minipage}}\par}} 

\def\keywords#1{{
	\centering{\begin{minipage}{4.5in}\baselineskip=10pt\footnotesize
	{\footnotesize\it Keywords}\/: #1
	\end{minipage}}\par}}

\newcommand{\bibit}{\nineit}
\newcommand{\bibbf}{\ninebf}
\renewenvironment{thebibliography}[1]
        {\frenchspacing
	 \ninerm\baselineskip=11pt
         \begin{list}{\arabic{enumi}.}
        {\usecounter{enumi}\setlength{\parsep}{0pt}     
	 \setlength{\leftmargin 12.7pt}{\rightmargin 0pt} 
         \setlength{\itemsep}{0pt} \settowidth
	{\labelwidth}{#1.}\sloppy}}{\end{list}}

\newcounter{itemlistc}
\newcounter{romanlistc}
\newcounter{alphlistc}
\newcounter{arabiclistc}

\newcommand{\fcaption}[1]{
        \refstepcounter{figure}
        \setbox\@tempboxa = \hbox{\footnotesize Fig.~\thefigure. #1}
        \ifdim \wd\@tempboxa > 5in
           {\begin{center}
        \parbox{5in}{\footnotesize\smalllineskip Fig.~\thefigure. #1}
            \end{center}}
        \else
             {\begin{center}
             {\footnotesize Fig.~\thefigure. #1}
              \end{center}}
        \fi}

\newcommand{\tcaption}[1]{
        \refstepcounter{table}
        \setbox\@tempboxa = \hbox{\footnotesize Table~\thetable. #1}
        \ifdim \wd\@tempboxa > 5in
           {\begin{center}
        \parbox{5in}{\footnotesize\smalllineskip Table~\thetable. #1}
            \end{center}}
        \else
             {\begin{center}
             {\footnotesize Table~\thetable. #1}
              \end{center}}
        \fi}

\def\@citex[#1]#2{\if@filesw\immediate\write\@auxout
	{\string\citation{#2}}\fi
\def\@citea{}\@cite{\@for\@citeb:=#2\do
	{\@citea\def\@citea{,}\@ifundefined
	{b@\@citeb}{{\bf ?}\@warning
	{Citation `\@citeb' on page \thepage \space undefined}}
	{\csname b@\@citeb\endcsname}}}{#1}}

\newif\if@cghi
\def\cite{\@cghitrue\@ifnextchar [{\@tempswatrue
	\@citex}{\@tempswafalse\@citex[]}}
\def\citelow{\@cghifalse\@ifnextchar [{\@tempswatrue
	\@citex}{\@tempswafalse\@citex[]}}
\def\@cite#1#2{{$\null^{#1}$\if@tempswa\typeout
	{IJCGA warning: optional citation argument 
	ignored: `#2'} \fi}}

\def\pmb#1{\setbox0=\hbox{#1}
	\kern-.025em\copy0\kern-\wd0
	\kern.05em\copy0\kern-\wd0
	\kern-.025em\raise.0433em\box0}

\def\fnt#1#2{\footnotetext{\kern-.3em
	{$^{\mbox{\scriptsize #1}}$}{#2}}}

\def\fpage#1{\begingroup
\voffset=.3in
\thispagestyle{empty}\begin{table}[b]\centerline{\footnotesize #1}
	\end{table}\endgroup}

\def\runninghead#1#2{\pagestyle{myheadings}
\markboth{{\protect\footnotesize\it{\quad #1}}\hfill}
{\hfill{\protect\footnotesize\it{#2\quad}}}}
\font\tenrm=cmr10
\font\tenit=cmti10 
\font\tenbf=cmbx10
\font\bfit=cmbxti10 at 10pt
\font\ninerm=cmr9
\font\nineit=cmti9
\font\ninebf=cmbx9
\font\eightrm=cmr8

\def\qed{\hbox{${\vcenter{\vbox{			
   \hrule height 0.4pt\hbox{\vrule width 0.4pt height 6pt
   \kern5pt\vrule width 0.4pt}\hrule height 0.4pt}}}$}}

\renewcommand{\thefootnote}{\fnsymbol{footnote}}	

\def\bsc{{\sc a\kern-6.4pt\sc a\kern-6.4pt\sc a}}  	
\def\bflatex{\bf L\kern-.30em\raise.3ex\hbox{\bsc}\kern-.14em 
T\kern-.1667em\lower.7ex\hbox{E}\kern-.125em X} 

\begin{document}

\runninghead
{Yu. Yu. Tarasevich, S. C. van der Marck}
{An Investigation of site-bond percolation}

\normalsize\textlineskip
\thispagestyle{empty}
\setcounter{page}{1}

\copyrightheading{Vol. 0, No. 0 (1999) 000--000}{}

\vspace*{0.88truein}

\fpage{1}
\centerline{\bf AN INVESTIGATION OF SITE-BOND PERCOLATION}
\vspace*{0.035truein}
\centerline{\bf ON MANY LATTICES}
\vspace*{0.37truein}
\centerline{\footnotesize YURIY YU. TARASEVICH}
\vspace*{0.015truein}
\centerline{\footnotesize\it 
Faculty of Physics and Mathematics,  Astrakhan State Pedagogical University}
\baselineskip=10pt
\centerline{\footnotesize\it   Astrakhan 414056, Russia}
\baselineskip=10pt
\centerline{\footnotesize\it   E-mail: vt@astranet.ru}
\baselineskip=10pt
\vspace*{10pt}
\centerline{\normalsize and}
\vspace*{10pt}
\centerline{\footnotesize STEVEN C. VAN DER MARCK\footnote{Present
address: EDS, P.O.~Box~406, 2260~AK~~Leidschendam, The~Netherlands}}
\vspace*{0.015truein}
\centerline{\footnotesize\it SIEP Research and Technical Services, P.O. Box~60}
\baselineskip=10pt
\centerline{\footnotesize\it  2280~AB~~Rijswijk, The~Netherlands}
\centerline{\footnotesize\it E-mail: steven.vandermarck@wxs.nl}
\vspace*{0.225truein}
\publisher{(received date)}{(revised date)}
\vspace*{0.21truein}
\abstracts{A calculation of site-bond percolation thresholds in
many lattices in two to five dimensions is presented.
The line of threshold values has been parametrized
in the literature, but we show here that there are
strong deviations from the known approximate equations.
We propose an alternative parametrization that lies
much closer to the numerical values.}{}{}

\vspace*{10pt}


\textheight=7.8truein
\setcounter{footnote}{0}
\renewcommand{\thefootnote}{\alph{footnote}}

\vspace*{1pt}\textlineskip	
\section{Introduction}	        
\vspace*{-0.5pt}
\noindent
\label{sec_intr}
Site-bond percolation is a natural generalization
of pure site percolation and pure bond percolation.
The generalization allows both sites and bonds
to be randomly occupied, with probabilities
$p_s$ and $p_b$ respectively, in the case of random
percolation.
In particular, if $p_b=p_s$,
site-bond percolation reduces to a pure
site or pure bond percolation on a decorated lattice.
It is also possible to introduce a correlation
between the occupation probabilities of adjacent
sites and bonds, which is usually referred to
as correlated percolation.
Site-bond percolation has many applications in 
different fields.
For example, it was used to describe the
sol-to-gel transition (gelation) of
polymers,\cite{coniglio}
the spreading of an infection,\cite{hammersley} etc. 

\begin{figure}[htbp]
\leavevmode  
   \epsffile{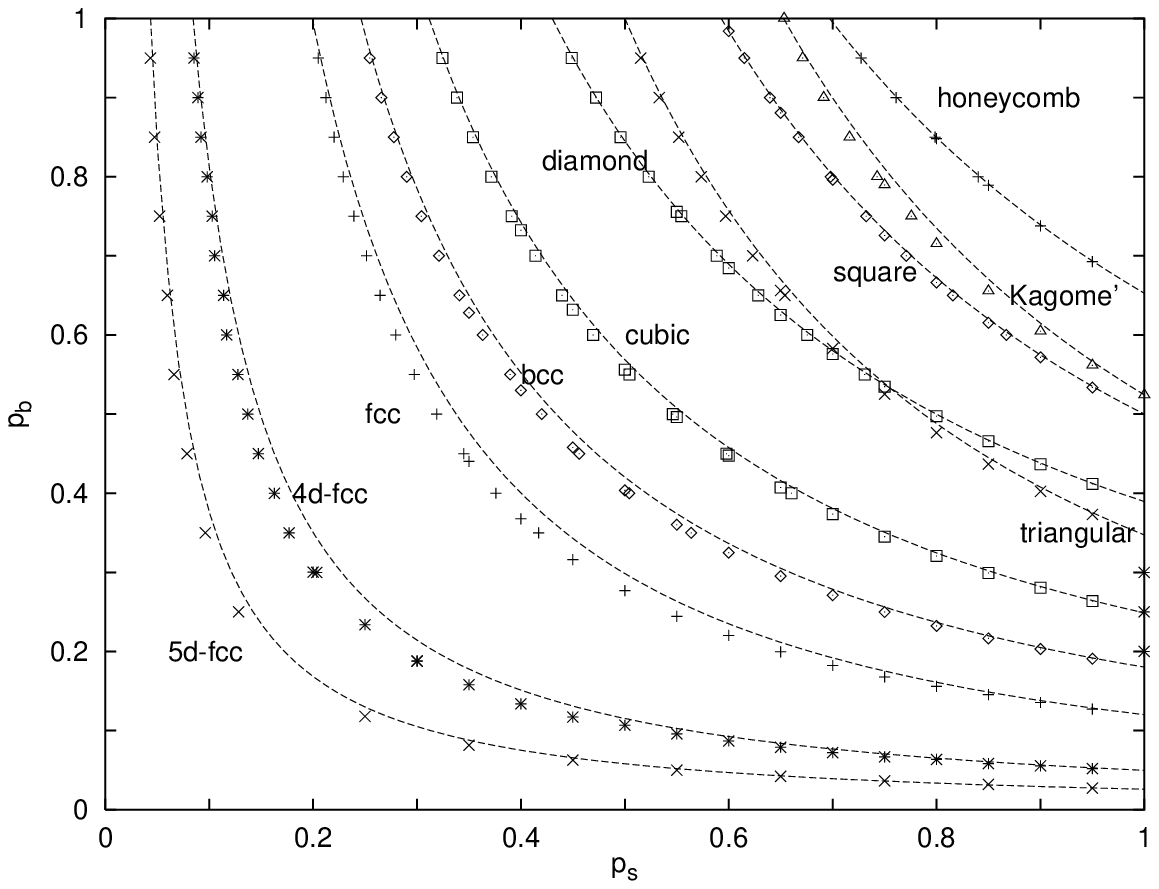}
  \fcaption{The percolation threshold pairs $(p_s,p_b)$ for
     many different lattices in various dimensions.
     The points are from the simulations
     (see Tables~\ref{tab_2d}--\ref{tab_5d}),
     the lines are from Eq.~(\ref{eq_YE}).  }
  \label{fig:YE}
\end{figure}

In site-bond percolation one looks for clusters
of occupied sites, connected to each other by
occupied bonds.
When there is a cluster that connects
one side of the system with the other,
the system is said to percolate.
First mentioned by Frisch and Hammersley,\cite{frisch}
the model was investigated at first by Hoshen
by means of Monte Carlo simulations.\cite{hoshen}
Agrawal et al.\cite{agrawal} showed, using a
series method, that the critical exponents of
pure site percolation are also valid for
site-bond percolation.
Nakanishi and Reynolds\cite{nakanishi} confirmed this
conclusion through a position-space renormalization group
study.
They also found that the boundary between percolation
and non-percolation in the $(p_b,p_s)$ plane is
governed by a single fixed point.
Hammersley\cite{hammersley} proved for a partially
directed graph a theorem.
The theorem yields the inequality
\begin{equation}
  \label{eq_hammer}
  P(p_{s}p_{b},1) \le P(p_{s},p_{b}) \le P(1,p_{s}p_{b}),
\end{equation}
where $P(p_{s},p_{b})$ is the percolation probability 
(i.e. the probability that a single source site is
connected to an infinite set of other sites).
The inequality Eq.~(\ref{eq_hammer}) gives
reasonably sharp bounds for the mixed percolation probability.
For a Bethe lattice ($p_{cb}=p_{cs}=p_c$, 
where $p_{cs}$ is the threshold for pure
site percolation, and $p_{cb}$ is the one for
pure bond percolation) we can get
\begin{equation}
\label{eq_bethe}
p_{s}p_b=p_c
\end{equation}
The shape of the boundary between percolation and
non-percolation was subsequently studied by
Yanuka and Englman.\cite{yanuka} They proposed the
following equation for the `critical curve'
in the $(p_b,p_s)$ plane.
\begin{equation}
   \frac{\log p_s}{\log p_{cs}} +
   \frac{\log p_b}{\log p_{cb}} = 1,
   \label{eq_YE}
\end{equation}
The equation satisfies
the inequality Eq.~(\ref{eq_hammer}) and
gives the correct limit Eq.~(\ref{eq_bethe})
for a Bethe lattice.
The equation defines a line in the $(p_b,p_s)$ plane,
starting at $(p_{cb},1)$ and ending at $(1,p_{cs})$,
as is shown in Figure~\ref{fig:YE}.
Yanuka and Englman showed plots of Eq.~(\ref{eq_YE})
together with simulation results for $(p_b,p_s)$
for the square, triangular, simple cubic and
face centered cubic (fcc) lattices.
All their simulation results lie remarkably close
to the line defined by Eq.~(\ref{eq_YE}).
It is not clear from their paper whether the data of
e.g. the fcc lattice, which lie visibly below the line
given by Eq.~(\ref{eq_YE}), are accurate enough to
conclude that there is a discrepancy.

Tretyakov and Inui\cite{tretyakov} studied the same
critical curve, but now for directed percolation.
Their simulation results were accurate enough to conclude
that deviations in the range $10^{-3}-10^{-4}$ occur.
This raises the question whether such deviations
also occur for `regular' site-bond percolation.

Therefore we have undertaken a systematic study
of the threshold values $(p_b,p_s)$ for
random site-bond percolation in many lattices.
The results, which are presented in
Section~\ref{sec_results},
show many deviations from the
Yanuka-Englman line of Eq.~(\ref{eq_YE}).
The deviations are largest for lattices
with a low ratio of $\alpha=p_{cb}/p_{cs}$,
i.e. lattices for which the bond percolation
thresholds is much lower than the site
percolation threshold.
In Section~\ref{sec_approx} we discuss
other approximative equations for the
critical line $(p_b,p_s)$.

\section{Numerical results}
\noindent
\label{sec_results}
\begin{table}[htbp]
  \begin{center}
  \tcaption{The results for the site-bond percolation
             thresholds for several lattices in two
             dimensions. Error estimates concerning the
             last digit(s) are indicated between brackets.
            }
  \label{tab_2d}
{\small
  \begin{tabular}{crlrlrlrl}
    \noalign{\smallskip}\hline\noalign{\smallskip}
        & \multicolumn{2}{c}{square} &
          \multicolumn{2}{c}{triangular} &
          \multicolumn{2}{c}{Kagom\'e} \\
    $p$ & $p_b(p_s\!=\!p)$ & $p_s(p_b\!=\!p)$ &
          $p_b(p_s\!=\!p)$ & $p_s(p_b\!=\!p)$ &
          $p_b(p_s\!=\!p)$ & $p_s(p_b\!=\!p)$ \\
    \hline\noalign{\smallskip}
    0.95&0.5334(3)&0.6149(3)&0.3732(3)&
         0.5152(4)&0.5615(4)&0.6711(4)\\
    0.90&0.5716(3)&0.6396(4)&0.4023(3)&
         0.5331(4)&0.6046(5)&0.6914(5)\\
    0.85&0.6154(3)&0.6673(3)&0.4368(3)&
         0.5518(4)&0.6556(6)&0.7162(5)\\
    0.80&0.6663(3)&0.6980(3)&0.4764(4)&
         0.5736(4)&0.7152(7)&0.7428(5)\\
    0.75&0.7257(4)&0.7322(4)&0.5251(4)&
         0.5969(4)&0.7894(9)&0.7757(8)\\
    0.70&0.7961(4)&0.7709(3)&0.5830(6)& 0.6231(3)\\
    0.65&0.8807(4)&0.8157(3)&0.6561(6)& 0.6541(4)\\
    0.60&0.9840(4)&0.8671(3)\\
    \noalign{\smallskip}
        & \multicolumn{2}{c}{octagonal} &
          \multicolumn{2}{c}{honeycomb} &
          \multicolumn{2}{c}{star-dual} \\
    $p$ & $p_b(p_s\!=\!p)$ & $p_s(p_b\!=\!p)$ &
          $p_b(p_s\!=\!p)$ & $p_s(p_b\!=\!p)$ &
          $p_b(p_s\!=\!p)$ & $p_s(p_b\!=\!p)$ \\
    \hline\noalign{\smallskip}
      0.95 & 0.3472(4) & 0.5133(7)& 0.6926(5)& 0.7275(5) &
             0.2796(4) & 0.5036(5) \\
      0.90 & 0.3744(5) & 0.5276(9)& 0.7377(5)& 0.7610(5) & 0.3022(5) & 0.5101(5) \\
      0.85 & 0.4087(5) & 0.5442(7)& 0.7890(5)& 0.7986(5) &
             0.3296(4) & 0.5188(5)\\
      0.80 & 0.4449(6) & 0.5620(7)& 0.8481(5)& 0.8401(5) & 0.3615(5) & 0.5285(5) \\
      0.75 & 0.4912(7) & 0.5816(7)&          &           &
             0.4006(4) & 0.5417(5)\\
      0.70 & 0.5481(10)& 0.6052(7)&          &           & 0.4499(5) & 0.5572(5) \\
      0.65 & 0.6213(24)&          &          &           & 0.5133(4) & 0.5763(5) \\
      0.60 &           &          &          &           & 0.5968(5)& 0.5984(5) \\

    \noalign{\smallskip}
        & \multicolumn{2}{c}{puzzle} &
          \multicolumn{2}{c}{direct} &
          \multicolumn{2}{c}{low-pb} \\
    $p$ & $p_b(p_s\!=\!p)$ & $p_s(p_b\!=\!p)$ &
          $p_b(p_s\!=\!p)$ & $p_s(p_b\!=\!p)$ &
          $p_b(p_s\!=\!p)$ & $p_s(p_b\!=\!p)$ \\
    \hline\noalign{\smallskip}
0.95& 0.5693(5) & 0.4495(4) & 0.5677(4) & 0.4433(4) & 0.2166(5)& 0.5003(5) \\
0.90& 0.5901(5) & 0.4849(5) & 0.5877(4) & 0.4780(4)\\
0.85& 0.6124(4) & 0.5247(5) & 0.6091(4) & 0.5183(5) & 0.2554(5)& 0.5048(5) \\
0.80& 0.6371(5) & 0.5720(5) & 0.6343(4) & 0.5648(4) \\
0.75& 0.6655(5) & 0.6283(5) & 0.6602(4) & 0.6199(5) & 0.3110(5)& 0.5156(5) \\
0.70& 0.6976(4) & 0.6952(5) & 0.6913(4) & 0.6876(5)  \\
0.65& 0.7328(5) & 0.7766(5) & 0.7263(4) & 0.7692(5) & 0.4020(5)& 0.5334(5) \\
0.60&           &           &0.7674(5)& 0.8712(7)     \\
0.55& & & & & 0.5644(5)\\
0.45& & & & & 0.6149(5)\\
0.35& & & & & 0.6997(6)\\
    \hline\noalign{\smallskip}
  \end{tabular}
}
  \end{center}
\end{table}

\begin{table}[htbp]
  \begin{center}
  \tcaption{The results for the site-bond percolation
             thresholds for several lattices in three
             dimensions. Error estimates concerning the
             last digit(s) are indicated between brackets.
            }
  \label{tab_3d}
  \begin{tabular}{crlrlrl}
    \noalign{\smallskip}\hline\noalign{\smallskip}
        & \multicolumn{2}{c}{sc}  &
          \multicolumn{2}{c}{fcc} &
          \multicolumn{2}{c}{hcp} \\
    $p$ & $p_b(p_s\!=\!p)$ & $p_s(p_b\!=\!p)$ &
          $p_b(p_s\!=\!p)$ & $p_s(p_b\!=\!p)$ &
          $p_b(p_s\!=\!p)$ & $p_s(p_b\!=\!p)$ \\
    \hline\noalign{\smallskip}
      0.95 & 0.2637(3)& 0.3244(3)& 0.1273(3)& 0.2049(5)
                                 & 0.1272(3)& 0.2052(3) \\
      0.90 & 0.2806(3)& 0.3385(3)& 0.1355(3)& 0.2124(4)
                                 & 0.1361(3)& 0.2125(3) \\
      0.85 & 0.2991(3)& 0.3539(3)& 0.1453(3)& 0.2202(4)
                                 & 0.1456(3)& 0.2214(3) \\
      0.80 & 0.3208(3)& 0.3716(3)& 0.1560(3)& 0.2291(5)
                                 & 0.1559(3)& 0.2306(3) \\
      0.75 & 0.3452(3)& 0.3912(3)& 0.1679(3)& 0.2391(5)
                                 & 0.1688(3)& 0.2404(3) \\
      0.70 & 0.3735(3)& 0.4139(3)& 0.1823(3)& 0.2515(5)
                                 & 0.1832(3)& 0.2519(3) \\
      0.65 & 0.4072(3)& 0.4395(3)& 0.1995(4)& 0.2644(5)
                                 & 0.2001(3)& 0.2653(3) \\
      0.60 & 0.4475(3)& 0.4694(3)& 0.2201(4)& 0.2796(6)
                                 & 0.2209(3)& 0.2806(3) \\
      0.55 &          &          & 0.2443(5)& 0.2974(6)
                                 & 0.2455(3)& 0.2985(3) \\
      0.50 &          &          & 0.2769(5)& 0.3190(6)
                                 & 0.2764(3)& 0.3195(3) \\
      0.45 &          &          & 0.3161(6)& 0.3451(6)
                                 & 0.3164(3)& 0.3454(3) \\
      0.40 &          &          & 0.3676(8)& 0.3759(6)
                                 & 0.3698(3)& 0.3764(3) \\
      0.35 &          &          &0.4404(11)& 0.4171(7) \\
    \noalign{\smallskip}
        & \multicolumn{2}{c}{bcc} &
          \multicolumn{2}{c}{diamond} \\
    $p$ & $p_b(p_s\!=\!p)$ & $p_s(p_b\!=\!p)$ &
          $p_b(p_s\!=\!p)$ & $p_s(p_b\!=\!p)$ \\
    \hline\noalign{\smallskip}
      0.95 & 0.1909(3) & 0.2545(5)& 0.4118(4)& 0.4489(5)\\
      0.90 & 0.2030(3) & 0.2656(5)& 0.4366(4)& 0.4723(5)\\
      0.85 & 0.2164(4) & 0.2777(5)& 0.4656(4)& 0.4959(5)\\
      0.80 & 0.2324(4) & 0.2900(5)& 0.4973(4)& 0.5229(5)\\
      0.75 & 0.2498(4) & 0.3043(5)& 0.5345(5)& 0.5544(6)\\
      0.70 & 0.2711(4) & 0.3210(6)& 0.5758(6)& 0.5885(5)\\
      0.65 & 0.2953(5) & 0.3408(6)& 0.6253(6)& 0.6283(5)\\
      0.60 & 0.3249(6) & 0.3633(6)& 0.6845(7)& 0.6755(6)\\
      0.55 & 0.3603(6) & 0.3896(7)& 0.7555(8)& 0.7308(6)\\
      0.50 & 0.4036(7) & 0.4199(7)\\
      0.45 & 0.4578(9) & 0.4562(7)\\
      0.40 & 0.5300(10)& 0.5043(7)\\
      0.35 & 0.6281(13)& 0.5639(7)\\
    \hline\noalign{\smallskip}
  \end{tabular}
  \end{center}
\end{table}

\begin{table}[htbp]
  \begin{center}
  \tcaption{The results for the site-bond percolation thresholds
    for the 4-dimensional simple cubic (sc) and fcc
    lattices. Error estimates concerning the last
    digit(s) are indicated between brackets.}
  \label{tab_4d}
  \begin{tabular}{crlrl}
    \noalign{\smallskip}\hline\noalign{\smallskip}
        & \multicolumn{2}{c}{4d-sc}
        & \multicolumn{2}{c}{4d-fcc} \\
    $p$ & $p_b(p_s=p)$ & $p_s(p_b=p)$
        & $p_b(p_s=p)$ & $p_s(p_b=p)$ \\
    \hline\noalign{\smallskip}
    0.95& 0.1701(6) & 0.2065(6) & 0.0520(03) & 0.0853(08)\\
    0.90&           &           & 0.0554(04) & 0.0892(09)\\
    0.85& 0.1900(6) & 0.2254(6) & 0.0584(04) & 0.0920(10)\\
    0.80&           &           & 0.0635(04) & 0.0979(09)\\
    0.75& 0.2183(6) & 0.2509(6) & 0.0668(05) & 0.1028(10)\\
    0.70&           &           & 0.0722(05) & 0.1052(11)\\
    0.65& 0.2547(6) & 0.2837(6) & 0.0789(06) & 0.1139(10)\\
    0.60&           &           & 0.0868(07) & 0.1167(13)\\
    0.55& 0.3064(6) & 0.3275(6) & 0.0957(07) & 0.1276(12)\\
    0.50&           &           & 0.1067(09) & 0.1372(13)\\
    0.45& 0.3834(6) & 0.3897(6) & 0.1171(10) & 0.1473(13)\\
    0.40&           &           & 0.1337(12) & 0.1627(15)\\
    0.35& 0.5083(6) & 0.4880(6) & 0.1579(14) & 0.1769(18)\\
    0.30&           &           & 0.1877(19) & 0.2034(17)\\
    0.25& 0.7512(6) & 0.6618(7) & 0.2339(27) \\
    \hline\noalign{\smallskip}
  \end{tabular}
  \end{center}
\end{table}

\begin{table}[htbp]
  \begin{center}
  \tcaption{The results for the site-bond percolation thresholds
     for the 5-dimensional fcc and Kagom\'e lattices.
     Error estimates concerning the last digit(s) are
     indicated between brackets.}
  \label{tab_5d}
  \begin{tabular}{crlrl}
    \noalign{\smallskip}\hline\noalign{\smallskip}
        & \multicolumn{2}{c}{5d-fcc}
        & \multicolumn{2}{c}{5d-Kagom\'e} \\
    $p$ & $p_b(p_s=p)$ & $p_s(p_b=p)$
        & $p_b(p_s=p)$ & $p_s(p_b=p)$ \\
    \hline\noalign{\smallskip}
    0.95& 0.0274(5) & 0.0434(5) & 0.1374(16) & 0.2173(20)\\
    0.85& 0.0319(5) & 0.0474(6) & 0.1555(17) & 0.2284(19)\\
    0.75& 0.0365(5) & 0.0522(6) & 0.1851(19) & 0.2439(20)\\
    0.65& 0.0424(5) & 0.0598(6) & 0.2111(19) & 0.2633(19)\\
    0.55& 0.0500(5) & 0.0661(6) & 0.2538(18) & 0.2957(19)\\
    0.45& 0.0627(6) & 0.0787(6) & 0.3245(20) & 0.3494(20)\\
    0.35& 0.0816(6) & 0.0963(6) & 0.4506(20) & 0.4243(19)\\
    0.25& 0.1179(6) & 0.1283(6) \\
    \hline\noalign{\smallskip}
  \end{tabular}
  \end{center}
\end{table}

\begin{figure}[htbp]
\leavevmode  
\epsffile{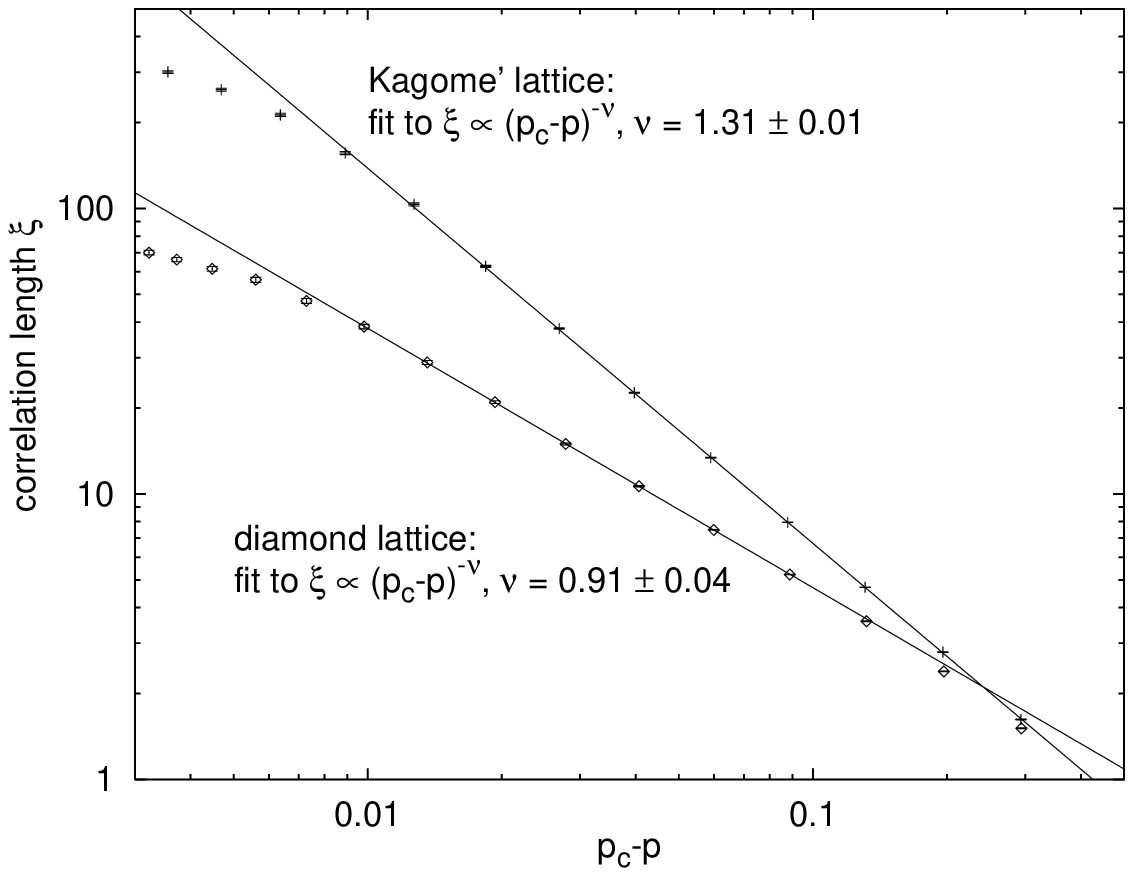}
  \fcaption{ The correlation length $\xi$ as a function of $p_c-p$.
     The runs were performed at
     $ \{p_b=0.65,\; p_s=0.6284(5)\} $
     for the diamond lattice and
     $ \{p_b=0.75,\; p_s=0.7757(8)\} $
     for the Kagom\'e lattice.  }
  \label{fig:ksi_dia}
\end{figure}

\begin{figure}[htbp]
\leavevmode  
\epsffile{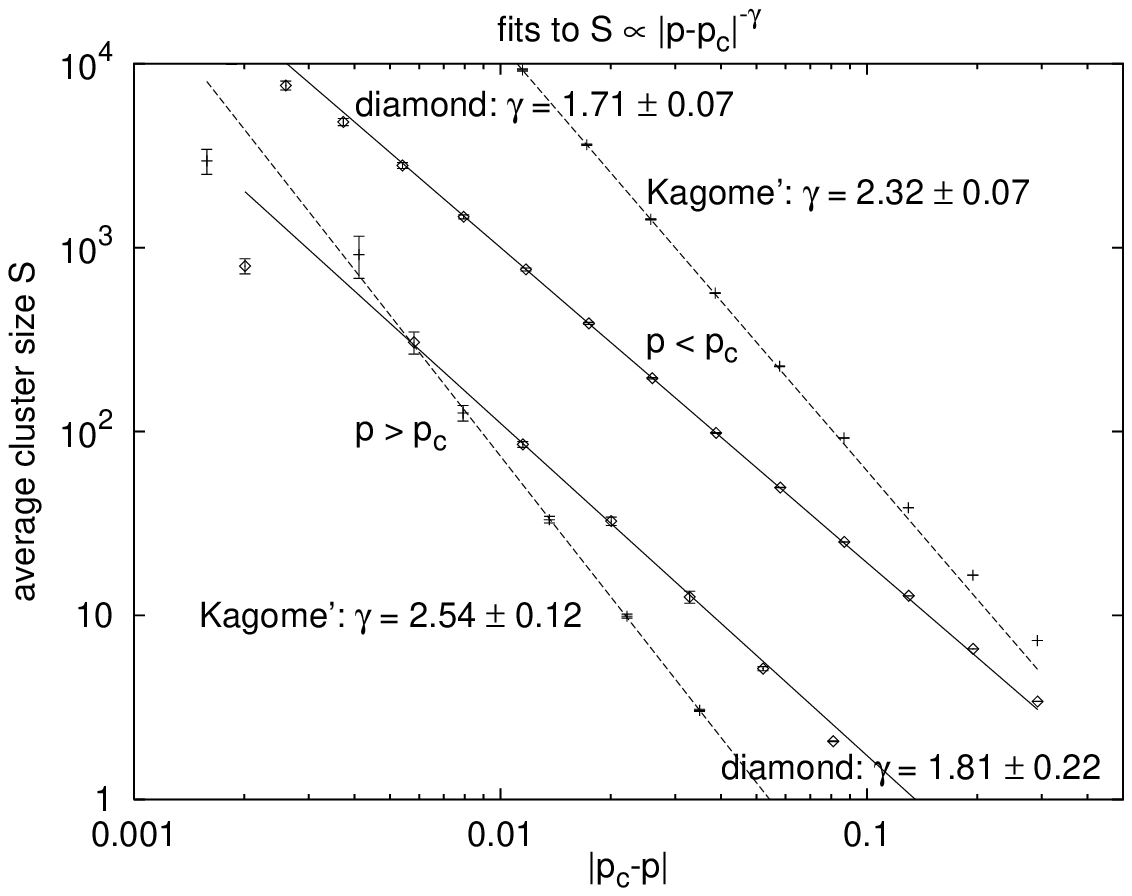}
  \fcaption{ The mean cluster size $S$ above and below the
     percolation threshold, as a function of $|p_c-p|$.
     The runs were performed at
     $ \{p_b=0.65,\; p_s=0.6284(5)\} $
     for the diamond lattice (solid lines) and
     $ \{p_b=0.75,\; p_s=0.7757(8)\} $
     for the Kagom\'e lattice (dashed lines).  }
  \label{fig:S_dia}
\end{figure}

\begin{figure}[htbp]
\leavevmode  
  \epsffile{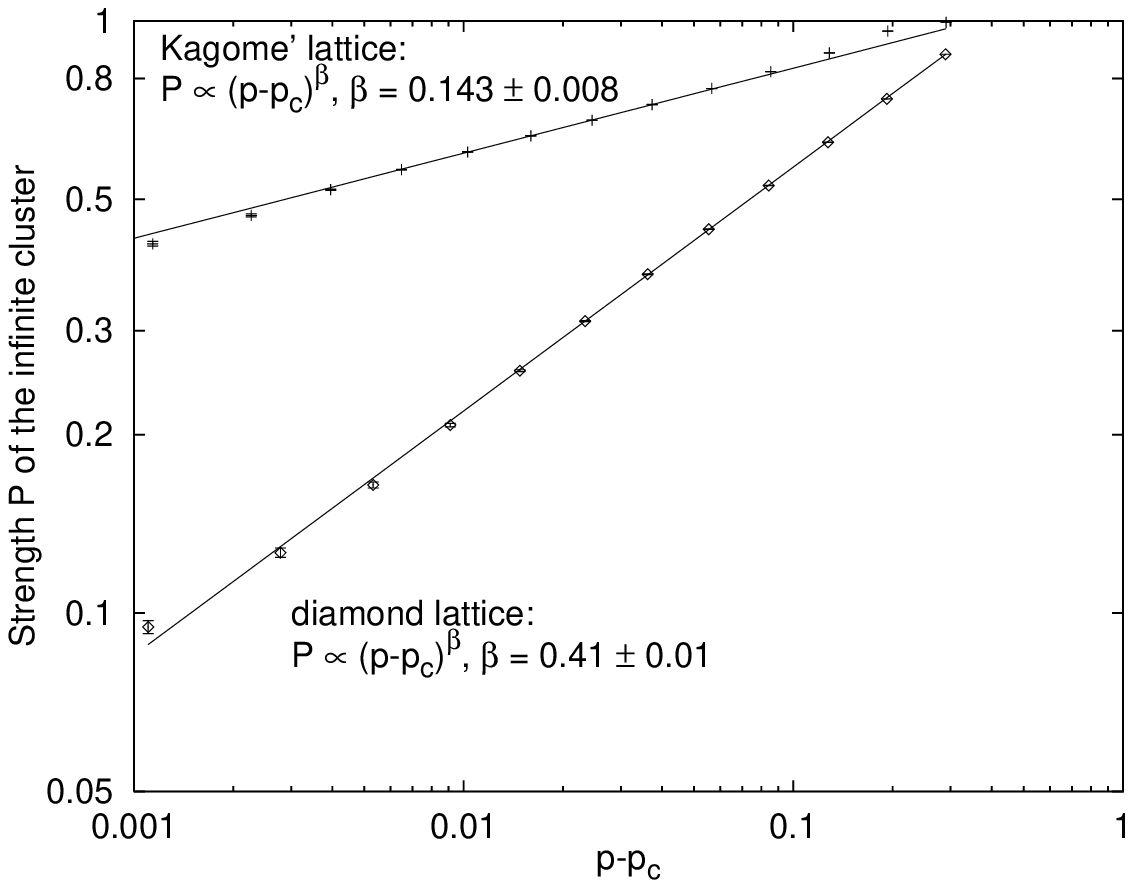}
  \fcaption{The strength P of the infinite cluster
     as a function of $p-p_c$.
     The runs were performed at
     $ \{p_b=0.65,\; p_s=0.6284(5)\} $
     for the diamond lattice and
     $ \{p_b=0.75,\; p_s=0.7757(8)\} $
     for the Kagom\'e lattice.  }
  \label{fig:P_dia}
\end{figure}

All numerical results were obtained with the
programs described in Ref.\cite{marck_98b}\footnote{The
programs can be obtained from one of the authors by
e-mail at steven.vandermarck@wxs.nl}.
For each lattice the percolation
threshold was calculated for at least three values of the
linear lattice size~$L$.
The percolation threshold $p_c(\infty)$
for infinite lattices can be
found by fitting these results for different lattice sizes
to the scaling relation
\begin{equation}
   \left| p_c(L) - p_c(\infty) \right|  \propto  L^{-1/\nu},
\end{equation}
where the critical exponent $\nu$ has the value
$4/3$ in two dimensions, and $0.88$, $0.68$, $0.57$
in 3, 4, and 5 dimensions respectively\cite{stauffer}.
These steps are precisely the same as for pure
site or pure bond percolation.
The results are shown in
Tables~\ref{tab_2d}--\ref{tab_5d} and in
Figures~\ref{fig:YE}--\ref{fig:yuriy}.
We have checked that for pure site percolation,
and for pure bond percolation, our program yields
results equivalent to more precise literature
values (see Table~\ref{tab_refs}).

We verified several scaling laws in percolation,
as a check on the program.
First, we calculated the
correlation length~$\xi$ below the percolation threshold,
as a function of $p_c-p$.
The correlation length scales as
\begin{equation}
   \xi \propto \left( p_c - p \right)^{-\nu}.
   \label{eq_scale_xi}
\end{equation}
The calculation was performed by first computing
the `gyration radius'~$R_s$ for a cluster of $s$~sites.
This radius is defined by
\begin{equation}
  R_s^2 = \frac{1}{s} \sum_{i=1}^{s}
               | {\bf r}_i - {\bf r}_0 |^2,
   \hspace*{1cm}
   \mbox{with}
   \hspace*{1cm}
  {\bf r}_o = \frac{1}{s} \sum_{i=1}^{s} {\bf r}_i,
\end{equation}
where ${\bf r}_i$ is the position of the $i^{th}$ site
in the cluster.
The correlation length~$\xi$ can then be calculated
as (see Ref.\cite{stauffer}, p. 50)
\begin{equation}
  \xi^2 = \frac{2\sum_s R_s^2 s^2 n_s}{\sum_s s^2 n_s},
\end{equation}
where $n_s$ is the mean number per lattice site
of clusters of $s$~sites.
We verified numerically that the correlation length
obeys the scaling relation~(\ref{eq_scale_xi}).
Our result for site-bond percolation on
the Kagom\'e and diamond lattices is shown
in Figure~\ref{fig:ksi_dia}.
A fit of the scaling relation
to the data yields $\nu = 1.31 \pm 0.01$ for the
Kagom\'e lattice, and $\nu = 0.91 \pm 0.04$ for the
diamond lattice.
This is within one or two standard deviations
of the established values of $\nu=4/3$ in two dimensions
and $\nu=0.88$ in three dimensions.

Also we have checked the scaling of the mean cluster
size~$S$ and the strength of the infinite cluster,
both as a function of~$|p-p_c|$.
The average cluster size~$S$ scales
with the same exponent both above and
below the percolation threshold:
\begin{equation}
   S \propto \left| p_c - p \right|^{-\gamma},
   \label{eq_scale_S}
\end{equation}
with $\gamma=43/18$ in two dimensions,
and $1.80$ in three\cite{stauffer}.
The results for site-bond percolation on
the Kagom\'e and diamond lattices
is shown in Fig.~\ref{fig:S_dia}.
The fits to the scaling relation~(\ref{eq_scale_S})
yield values for~$\gamma$ that are consistent with the
literature values.

The strength~$P$ of the infinite cluster scales as
\begin{equation}
   P \propto \left( p - p_c \right)^{\beta},
   \label{eq_scale_P}
\end{equation}
with $\beta=5/36$ in two dimensions,
and $0.41$ in three\cite{stauffer}.
The results for site-bond percolation on
the Kagom\'e and diamond lattices
is shown in Fig.~\ref{fig:P_dia}.
The fits to the scaling relation~(\ref{eq_scale_P})
yield values for~$\beta$ that are consistent with the
literature values.

\section{Approximate equations}
\begin{figure}[htbp]
\leavevmode  
\epsfxsize=12.5cm
\epsffile{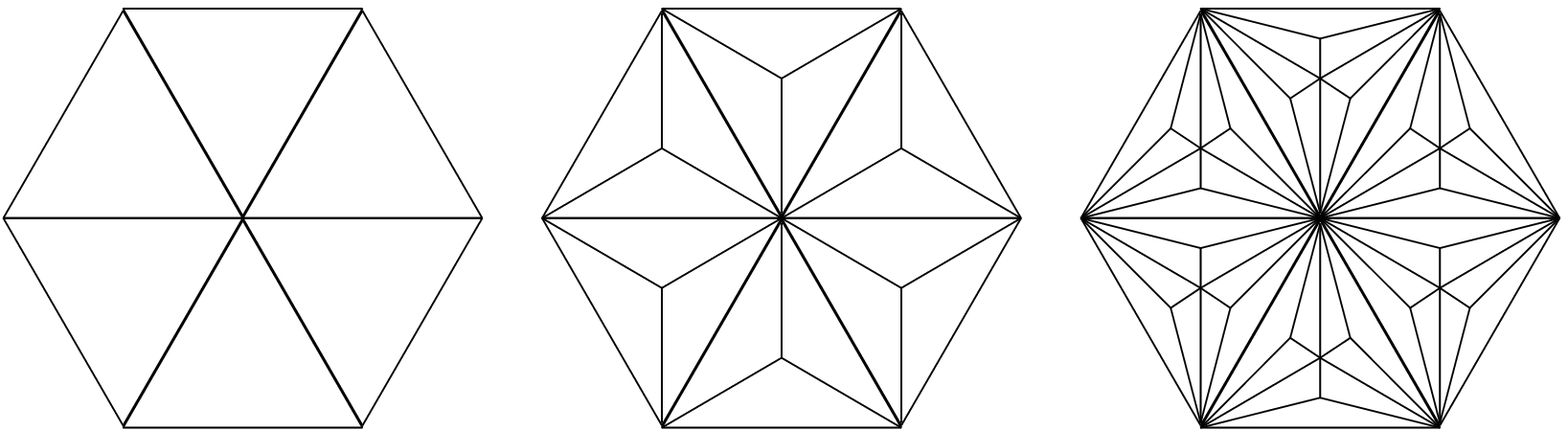}
\smallskip
  \fcaption{The triangular lattice (left) can be
             modified by adding a site in the middle of each
             triangle (middle). The same operation once more
             yields the lattice on the right.
             All these lattice have $p_s=\frac{1}{2}$,
             but the bond percolation threshold decreases
             from left to right.}
  \label{fig:lowpb}
\end{figure}

\noindent
\label{sec_approx}
It is clear from Figure~\ref{fig:YE} that there are
systematic deviations from the Yanuka-Englman line.
The accuracy of the numerical values, as indicated
in Tables~\ref{tab_2d}--\ref{tab_5d} is easily
high enough to warrant this conclusion. The Yanuka-Englman line is 
situated above the numerical values for all calculated examples. One can say it 
is an upper estimation for the real critical curve. 
In general it seems that the deviations are
largest for lattices
with a low ratio of $\alpha=p_{cb}/p_{cs}$,
i.e. lattices for which the bond percolation
thresholds is much lower than the site.
The most clear example is the `low-$p_b$' lattice
(see Fig.~\ref{fig:lowpb}), which has $p_{cs}=0.5$
and $p_{cb}=0.2006(3)$ (see Table~\ref{tab_refs}).
This lattice has been constructed from the
triangular lattice in two steps,
see Fig.~\ref{fig:lowpb}. All the lattices in
this figure are fully triangulated and have
therefore $p_{cs}=\frac{1}{2}$, as shown by Sykes
and Essam\cite{sykes}. The bond percolation
threshold, however, will decrease with every step
in Fig.~\ref{fig:lowpb}, and these steps can be
repeated even further. In this way one can construct
lattices with arbitrarily low ratio $p_{cb}/p_{cs}$.

One of the most remarkable results in our calculations is 
the results for two pairs of lattices with very close or
equal critical probabilities. One pair is the face centered
cubic (fcc) lattice and the hexagonal close packed (hcp)
lattice. Another pair, in two dimensions, is the direct
lattice and the puzzle lattice (we use here the nicknames
as in Ref.~\cite{ziff_99}).
The fcc and hcp lattices, although they are topologically different,
have $p_{cb}^{fcc}=p_{cb}^{hcp}$
and  $p_{cs}^{fcc}=p_{cs}^{hcp}$
within the present small error margins (see Table~\ref{tab_refs}). 
Also the critical curves we calculated for these lattices
are the same within our error margin. 
In two dimensions, the pure site and pure bond percolation thresholds
of the direct and the puzzle lattice are almost the same.
Also for this pair the critical curves lie close together.

These results suggest that once the pure site and pure bond thresholds
are known, the whole critical curve is fixed.
In other words, there is no need to use extra parameters (such as
critical exponents, dimension, or coordination number)
to describe a critical curve but $p_{cb}$ and $p_{cs}$. 
In this context it is interesting to note that in
the limiting case of Bethe lattices, Eq.~(\ref{eq_bethe}),
there are also no parameters other than~$p_c$.

One can try to find an equation for the critical curve
by making use of the work of Hammersley,
who derived boundaries 
for mixed percolation\cite{hammersley}
$$
   p_{s}^{up}(p_{b})=\frac{p_{cs}}{p_{b}},
$$
and
$$
   p_{s}^{down}(p_{b})=\frac{p_{cb}}{p_{b}}.
$$
Let us look for a critical relationship
as a combination of these boundaries
$$
   p_{s}(p_{b}) = C_{1}(p_b) \cdot p_{s}^{down}(p_{b}) +
                  C_{2}(p_b) \cdot p_{s}^{up}  (p_{b}).
$$
Take the functions $C_{1}(p_b)$ and $C_{2}(p_b)$ in the simplest form
$$
   C_{1}(p_b) = \frac{1-p_b}{1-p_{cb}}
              = \left\{
                   \begin{array}{ll}
                       1 & \mbox{for } p_b=p_{cb} \\
                       0 & \mbox{for } p_b=1
                   \end{array}
                \right. ,
$$
and
$$
   C_{2}(p_b) = \frac{p_b-p_{cb}}{1-p_{cb}}
              = \left\{
                   \begin{array}{ll}
                       1 & \mbox{for }  p_b=1 \\
                       0 & \mbox{for }  p_b=p_{cb}
                   \end{array}
                 \right. .
$$
After some transformation we have 
\begin{equation}
   p_b \left( p_s + A \right) = B,
   \label{eq_yuriy}
\end{equation}
where
\begin{eqnarray}
     A & = &   \frac{p_{cb}-p_{cs}}{1-p_{cb}}, \\
     B & = & p_{cb} \frac{1-p_{cs}}{1-p_{cb}}.
\end{eqnarray}

\begin{figure}[htbp]
\leavevmode  
      \epsffile{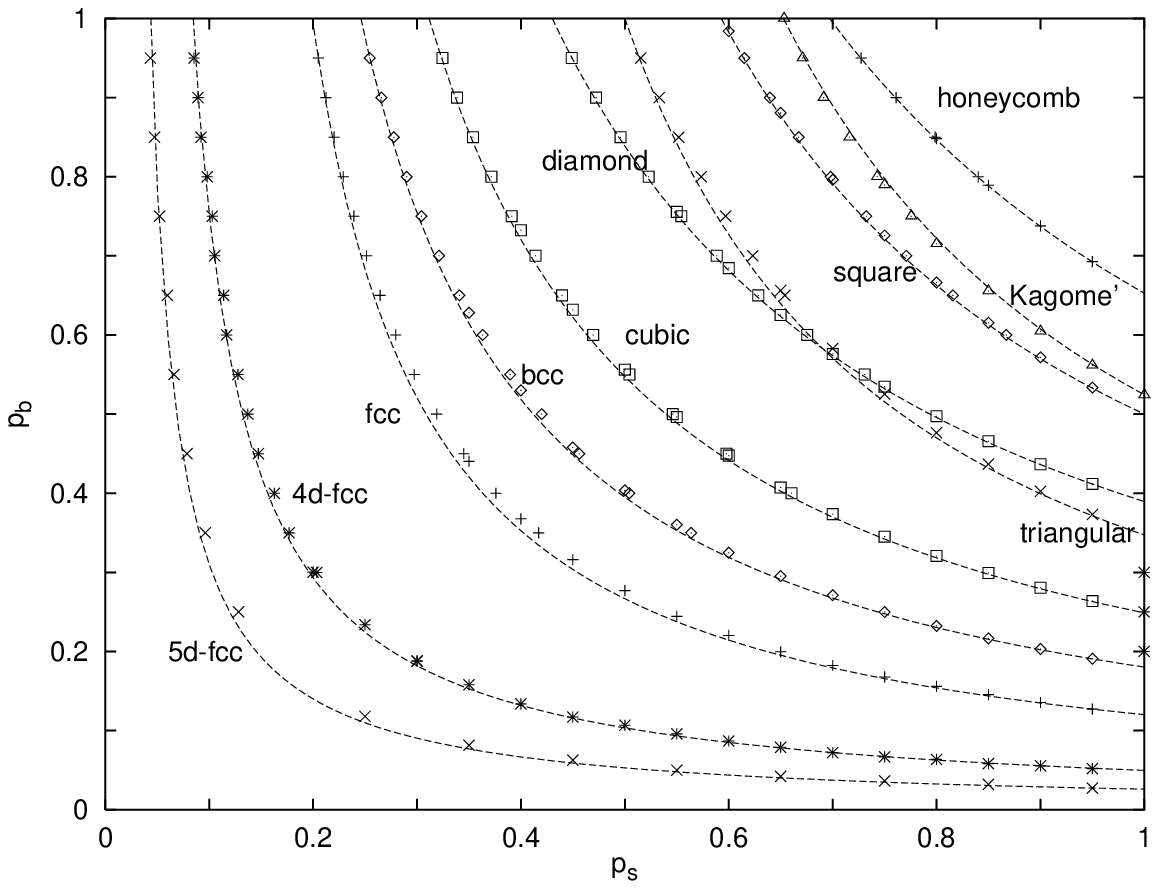}
  \fcaption{The percolation threshold pairs $(p_s,p_b)$ for
     many different lattices in various dimensions.
     The points are from the simulations
     (see Tables~\ref{tab_2d}--\ref{tab_5d}),
     the lines are from Eq.~(\ref{eq_yuriy}).
  }
  \label{fig:yuriy}
\end{figure}

The relationship Eq.~(\ref{eq_yuriy}) gives almost the same results 
as Eq.~(\ref{eq_YE}) provided $p_b$ and $p_s$ are close
(square, simple cubic, octagonal, etc.).
In these cases both curves and the numerical points are in good
agreement with one another.
In other cases Eq.~(\ref{eq_yuriy}) gives much better approximations
than Eq.~(\ref{eq_YE}) (see Figure~\ref{fig:yuriy}).

We examined a number of different relationships with 2 and 3 parameters  
to describe 
the numerical results. Although the results were excellent for some lattices,
we could not find a relationship suitable for all lattices.

Hence the problem to find the real law for critical probabilities 
is still unsolved. 
We hope this work with the large collection of high precision 
numerical results and the new
estimation of critical probabilities will assist to solve the problem.

\nonumsection{Acknowledgements}
\noindent
One of the authors (Yu.Yu. Tarasevich) thanks Max-Plank-Institut 
f\"ur Physik Komplexer Systeme for the possibility to work
in the library during a seminar and Prof. Amnon Aharony
for discussions.

\begin{table}[htbp]
  \begin{center}
  \tcaption{The site and bond percolation thresholds
      used in the article,
      together with the references to the literature.}
  \label{tab_refs}
  \begin{tabular}{lllll}
    \noalign{\smallskip}\hline\noalign{\smallskip}
       lattice & \multicolumn{2}{c}{$p_s$ [Ref.]}
               & \multicolumn{2}{c}{$p_b$ [Ref.]} \\
    \hline\noalign{\smallskip}
  honeycomb & 0.697\,043\,(2)      & \cite{ziff_99}
            & 0.652\,703\,$\ldots$ & \cite{sykes}       \\
  Kagom\'e  & 0.652\,703\,$\ldots$ & \cite{sykes}
            & 0.524\,405\,3(3)     & \cite{ziff_97}     \\
  square    & 0.592\,746\,0(5)     & \cite{ziff_86}
            & 0.5                  & \cite{kesten}      \\
  triangular& 0.5                  & \cite{sykes,kesten}
            & 0.347\,296\,$\ldots$ & \cite{sykes}       \\
  octagonal & 0.5                  & \cite{sykes}
            & 0.3230(2)      & \cite{marck_97a,marck_99}\\
  star-dual & 0.5                  & \cite{sykes}
            & 0.2593(2)            & \cite{marck_99}    \\
  low-$p_b$ & 0.5                  & \cite{sykes}
            & 0.2006(3)            & \\
  puzzle    & 0.550806(2)          & \cite{ziff_99}
            & 0.4142(2)            &\\
  direct    & 0.550213(2)          & \cite{ziff_99}
            & 0.4195(2)\\

  \noalign{\smallskip}
  simple cubic & 0.311\,608\,0(3)  & \cite{lorenz_98b}
            & 0.248\,812\,6(5)     & \cite{lorenz_98a}  \\
  bcc       & 0.245\,961\,5(10)    & \cite{lorenz_98b}
            & 0.180\,287\,5(10)    & \cite{lorenz_98a}  \\
  fcc       & 0.199\,236\,5(10)    & \cite{lorenz_98b}
            & 0.120\,163\,5(10)    & \cite{lorenz_98a}  \\
  hcp       & 0.199\,24(5)         & \cite{marck_97a}
            & 0.120\,15(5)         & \cite{marck_97a} \\
  diamond   & 0.4301(2)            & \cite{marck_98b}
            & 0.3893(2)            & \cite{marck_97a}   \\
  \noalign{\smallskip}
  4d-cubic  & 0.196901(5)          & \cite{balles97}
            & 0.1600(2)            & \cite{gaunt_78}    \\
  4d-fcc    & 0.0842(3)            & \cite{marck_98b}
            & 0.049(1)             & \cite{marck_98b}   \\
  \noalign{\smallskip}
  5d-fcc    & 0.0431(3)            & \cite{marck_98b}
            & 0.026(2)             & \cite{marck_98b}   \\
 5d-Kagom\'e& 0.2084(4)            & \cite{marck_98a}
            & 0.130(2)             & \cite{marck_98b}   \\
    \noalign{\smallskip}
    \hline\noalign{\smallskip}
  \end{tabular}
  \end{center}
\end{table}

\nonumsection{References}

\end{document}